\documentstyle[prb,aps,multicol]{revtex}
\renewcommand{\narrowtext}{\begin{multicols}{2} \global\columnwidth20.5pc}
\renewcommand{\widetext}{\end{multicols} \global\columnwidth42.5pc}
\begin{document}
\draft
\title{
Understanding the dynamics of fractional edge states with composite fermions}
\author{ Dmitri B. Chklovskii and Bertrand I. Halperin}
\address{
Physics Department, Harvard University, Cambridge, MA 02138}
\maketitle

\begin{abstract}
Fractional edge states can be viewed as integer edge states of composite
fermions. We exploit this to discuss
 the conductance of the fractional quantized Hall states and the
velocity of edge magnetoplasmons.

\end{abstract}

\narrowtext

One of the most interesting developments in the physics of the quantized Hall
effect was the introduction of the composite fermion
 approach.\cite{Jain,Lopez,Kalmeyer,HLR}
It is realized through an exact transformation of the initial
electron system to the system of composite fermions by attaching two flux
quanta to each particle. The mean-field approximation to the new system
already contains much of the physics related to electron-electron correlations.
In particular it allows one to understand fractional quantized Hall states as 
integer quantized Hall states of composite fermions. Because of the presence
of an excitation gap in the bulk, edge states play an important role in these
systems. It is natural then to apply composite fermions to study fractional
edge states. This has been done using the Hartree approximation for composite
fermions\cite{Brey,Chklovskii}, which yields ground state properties such as 
the density distribution 
and the number of edge channels for various filling factors. One has to go 
beyond the Hartree approximation in order to study the dynamics of the
edge states, which involves excited states.

A significant progress in the understanding of edge states dynamics has been
 achieved by using the method of chiral bosonization\cite{Wen}. However, this
method relies on postulating the number and the direction of propagation of
edge modes. The composite fermion approach allows one to derive
 this quantities microscopically. Once the velocities and interactions of edge
 modes are known they can be used as input parameters in the method of chiral
bosonization.

In this paper we use the RPA to study fractional edge states. We begin
by reviewing ``gauge argument'' and applying it to the system of
 composite fermions. This leads us to the introduction of the composite
fermion electrochemical potentials which are very useful in calculating
conductance of the system. 

 Following Laughlin\cite{Laughlin} and Halperin\cite{Halperin82} we consider
 an idealized clean
2DEG sample of the annular shape at filling factor $\nu$. As the magnetic flux
 threading the hole is increased adiabatically by one flux quantum ($\phi_0$)
 $\nu$ electrons are transferred from the outer to the inner edge of the
 sample.
This leads to the total energy change given by $\nu(\xi_o-\xi_i)$, where
 $\xi_o-\xi_i$ is the difference between electrochemical potentials on the 
two edges. Since the work done on the system is given by $I\phi_0/c$, where 
$I$ is the current around the annulus,
\begin{equation}
\label{cur}
I=\nu(e/h)(\xi_o-\xi_i).
\end{equation}
Now we rederive this result by using composite fermions. The electron system 
at filling factor $\nu=p/(2p+1)$ is viewed as the system of composite fermions
at integer filling factor $p$. Naively, increasing the magnetic flux threading
the annulus by one flux
 quantum will transfer $p$ composite fermions between the outer and the inner
edge. However the change in the flux $\Delta \phi$ enclosed by the annulus
 median comes from the flux carried by composite fermions as well as from the
 external source. Thus $\Delta \phi=(1-2x)\phi_0$, where $x$ is the net
 number of
  composite fermions transferred between the edges. By relating the net 
transferred charge to the change in flux, using
 $x=p\Delta \phi/\phi_0$, we find that $x=p/(2p+1)$ in agreement with the
 electron result.

It is helpful to introduce the composite fermion electrochemical potential, 
$\xi^f$. We
 define it as a change in the total energy resulting from adiabatically 
adding a composite fermion to the edge. The difference from the electron
potential becomes clear in view of the previous discussion. The addition
of a composite fermion to the edge involves turning on of the Chern-Simons
flux, which adds or removes electrons from the edge. From the gauge argument
we find the following relation between the two potentials:
\begin{equation}
\label{ppot}
\xi^f=\xi/(2p+1)
\end{equation}
This relation has been derived in a different way by Kirczenow and Johnson.\cite{kir}

If the edges of the annulus are sufficiently well separated 
currents associated with each edge can be defined. These currents must depend
only on the electrochemical potential of the corresponding edge. Since the
total current is given by Eq.\ref{cur} the edge current is
\begin{equation}
\label{b}
I=\nu(e/h)\xi.
\end{equation}
 Expressing this result in terms of the composite fermion potential 
\begin{equation}
\label{bare}
I=p(e/h)\xi^f.
\end{equation}
For the case $|p|>1$, there will be several states at each edge, and in the 
ideal case one can define separate chemical potentials $\xi^f$ on each edge 
state. We have assumed that there is a sufficient amount of scattering due
to impurities or phonons, so that the states at a given edge are
equilibrated, and there is a single chemical potential at each edge.

Now we consider an electromagnetic response of a
 single edge at a finite wavevector along the edge.
The single edge approximation is justified if the wavelength is smaller
 than the distance between the edges. For a given bulk filling factor there can
 be several magnetoplasmon modes at the edge if the impurity scattering is
not too great. Here we restrict ourselves
to the case of a sharp confining potential. In this case
there is only one mode present for simple fractions $\nu=1/(2k+1)$. 
We focus on the $\nu=1$ case where the composite fermion result can be 
easily compared against the electron calculation. We can use previously derived
formulas setting $p=-1$. Our goal
is to find a pole in the response function and identify it with the
magnetoplasmon mode. 

We start by deriving the electron result. For frequencies smaller than the 
cyclotron frequency and wavelengths
larger than the magnetic length eq.\ref{b} can be written in terms
of Fourier harmonics
\begin{equation}
\label{bb}
I(k,\omega)=(e/h)\xi(k,\omega).
\end{equation}
The electrochemical potential can be represented 
as a sum of the external potential and the induced potential
\begin{equation}
\label{sta}
\xi(k,\omega)=\xi^e(k,\omega)+\xi^i(k,\omega)
\end{equation}
The induced electrochemical potential is proportional to the electron density
 at the edge
\begin{equation}
\xi^i(k,\omega)=n(k,\omega)\partial\xi/\partial n.
\end{equation}
The coefficient $\partial\xi/\partial n$ is independent of $k$ in the limit
  $k\rightarrow 0$, if the electron electron interaction is short ranged; for
unscreened Coulomb interactions we have 
$\partial\xi/\partial n \sim |{\rm ln}k|$.\cite{VM}
For the wavelengths smaller than the length of the edge we invoke the
 continuity equation $n(k,\omega)=I(k,\omega)k/\omega$ to get
\begin{equation}
\xi^i(k,\omega)=I(k,\omega)(k/\omega)\partial\xi/\partial n 
\end{equation}
By substituting $I(k,\omega)$ from the eq.\ref{bb} and solving for $\xi(k,\omega)$ we find
\begin{equation}
\label{frr}
\xi(k,\omega)=\frac{\xi^e(k,\omega)}{1-(k/\omega)(e/h)\partial\xi/\partial n}
\end{equation}
Substituting this expression in eq.\ref{bb} we obtain the full response 
function
\begin{equation}
\label{fin}
I(k,\omega)=\frac{(e/h)\xi^e(k,\omega)}{1-(k/\omega)(e/h)\partial\xi/\partial n}
\end{equation}

 At finite $k$ and $\omega$ there is a pole in the response
function, which can be identified with the magnetoplasmon mode. Because of the
stability considerations $\partial\xi/\partial n$ has to be positive. Then  
the direction of propagation is  in agreement with the 
classical magnetoplasmon result.\cite{VM}

Attachment of two flux quanta transforms the $\nu=1$ state into the $\nu=-1$
 state with the effective magnetic field reversed. Therefore, on the mean 
field level one might  expect that the composite fermion approach gives
the reverse direction of propagation for the magnetoplasmon mode.
In fact this is the result that comes out of the Hartree approximation,
 where  the induced Chern-Simons field is neglected.
Let us demonstrate that including properly the Chern-Simons electric field
gives the expected direction of propagation in the composite fermion approach.
Eq.\ref{bare} can be written in terms of Fourier harmonics
\begin{equation}
\label{fb}
I(k,\omega)=-(e/h)\xi^f(k,\omega).
\end{equation}
The composite fermion electrochemical potential $\xi^f$ may be related to the 
electron potential $\xi$ and the edge current $I$ by
\begin{equation}
\label{chh}
\xi(k,\omega)=\xi^f(k,\omega)+2(h/e)I(k,\omega)
\end{equation}
where the second term results from the Chern-Simons electric field produced by
the current. [ Eqs.\ref{fb} and \ref{chh} imply Eq.\ref{ppot}, as required ]
 Substituting Eq. \ref{chh} in Eq. \ref{sta} we find
\begin{equation}
\xi^f(k,\omega)=\xi^e(k,\omega)-2I(k,\omega)h/e+n(k,\omega)\partial\xi/\partial n
\end{equation}
Using the continuity equation and expressing current in terms of the 
composite fermion electrochemical potential we get
\begin{equation}
\xi^f(k,\omega)=\xi^e(k,\omega)+2\xi^f(k,\omega)-(k/\omega)\xi^f\partial\xi/\partial n 
\end{equation}
Solving this equation for $\xi^f$ we find
\begin{equation}
\xi^f(k,\omega)=\frac{\xi^e(k,\omega)}{-1+(k/\omega)(e/h)\partial\xi/\partial n}
\end{equation}
Both of the terms in the denominator change signs here compared to Eq.\ref{frr}
thus giving the right direction of propagation for the magnetoplasmon mode.
In Hartree approximation the Chern-Simons contribution to the electrochemical
 potential $2I(k,\omega)h/e$  is ignored thus yielding an incorrect direction
of propagation. 

For more complicated filling fractions $\nu$, where there are two or more edge
 states at a single edge, it is necessary to take into account interactions
between charge fluctuations associated with the different edge states.
 Similarly, if one considers wavelengths which are not large compared to the
magnetic length and/or the geometric width of the edge, it is
necessary to take into account more details of the electron wavefunctions.
A natural approximation to use in this case is the RPA (or time dependent
Hartree approximation) where the composite fermions are treated as 
non-interacting particles, driven by the space and time-dependent scalar and
 vector potentials. For a single edge state, the integral equation which
results from the RPA reduces properly to the above equations in the limit
where $k$ and $\omega$ are small, giving the correct direction of propagation
as expected.

We are grateful to A. Stern for helpful discussions. This work has been
supported by the Harvard Society of Fellows and by NSF Grant No. DMR94-16910.

\widetext

\begin{thebibliography}{99}
\bibitem{Jain} J. K. Jain, Phys.\ Rev.\ Lett.\  {\bf 63}, 199 (1989).
\bibitem{Lopez} A. Lopez and E. Fradkin, Phys. Rev. B, {\bf 44}, 5246 (1991).
\bibitem{Kalmeyer} V. Kalmeyer and S.-C. Zhang, Phys. Rev. B {\bf 46}, 9889
(1992).
\bibitem{HLR} B. I. Halperin, P. A. Lee, and N. Read, Phys.\ Rev.\ B\ {\bf 47},
7312 (1993).
\bibitem{Brey} L. Brey, Phys. Rev. B {\bf 50}, 11861 (1994).
\bibitem{Chklovskii} D. B. Chklovskii, Phys. Rev. B {\bf 51}, 9895 (1995).
\bibitem{Wen} X.-G. Wen, J. Mod. Phys. B {\bf6}, 1711 (1992).
\bibitem{Laughlin} R. B. Laughlin, Phys. Rev. B {\bf 23}, 5632 (1981).
\bibitem{Halperin82} B.I. Halperin, Phys. Rev. B  {\bf 25}, 2185 (1982).
\bibitem{kir} G. Kirczenow and B.L. Johnson, Phys. Rev. B {\bf 51}, 17579 (1995).
\bibitem{VM} V.A.Volkov, S.A. Mikhailov, Sov. Phys. JETP {\bf 67}, 1639 (1988).
\end{thebibliography}
\end{document}